\documentclass{llncs} 
\usepackage[dvips]{graphicx}
\usepackage{llncsdoc}
\usepackage{tabularx}
\usepackage{amssymb}
\usepackage{amsmath}
\usepackage{longtable}

\newcommand{\bd}[1]{\mbox{\boldmath $#1$}} 
\newcommand{\cc}{\mathcal N}
\newcommand{\ny}{\left\{}
\newcommand{\zr}{\right\}}

\newcommand{\RR}{\mathcal R}
\newcommand{\argmax}[1]
           { {\renewcommand{\arraystretch}{0.5}
              \begin{array}{ccc} ~ \\ \mbox{\rm argmax} \\ {\scriptstyle #1}
              \end{array} } }

\begin{document}
\title{On~class~visualisation~for~high~dimensional~data:\\Exploring~scientific~data~sets}
\author{Ata Kab\'{a}n${}^1$, Jianyong Sun${}^{1,2}$, Somak Raychaudhury${}^2$ and Louisa Nolan${}^2$}
\institute{${}^1$ School of Computer Science \hspace{0.5cm}  ${}^2$ School of Physics and Astronomy\\ 
The University of Birmingham, Birmingham, B15 2TT, UK\\
\email{\{axk,jxs\}@cs.bham.ac.uk, \{somak,lan\}@star.sr.bham.ac.uk}}

%

\maketitle

\begin{abstract}
Parametric Embedding (PE) has recently been proposed as a general-purpose algorithm for class visualisation. It takes class posteriors 
produced by a mixture-based clustering algorithm and projects them in 2D for visualisation. However, although this fully modularised combination of objectives 
(clustering and projection) is attractive for its conceptual simplicity, in the case of high dimensional data, we show that a more optimal combination of these objectives can be achieved by 
integrating them both into a consistent probabilistic model. In this way, the projection step will fulfil a role of regularisation, 
guarding against the curse of dimensionality. 
As a result, the tradeoff between clustering and visualisation turns out to enhance the predictive abilities of the overall model.
We present results on both synthetic data and two real-world high-dimensional data sets: observed spectra of early-type galaxies and gene expression arrays.
\end{abstract}
\section{Introduction}
Clustering and visualisation are two widespread unsupervised data analysis techniques, with applications in numerous fields of 
science and engineering. Two key strategies can be distinguished in the 
literature: (1) One is to produce a compression
of the data first and then use that to visually detect distinct clusters. A wide range of linear and nonlinear dimensionality reduction techniques developed in machine learning 
follow this route, including PCA, GTM \cite{GTM}, etc.
(2) The alternative strategy is to cluster the data first and visualise the resulting cluster assignments afterwards. This is 
more popular in the data mining community \cite{DM}.
A recently proposed method, termed Parametric Embedding (PE) \cite{PE} proposes to take class posteriors 
produced by a mixture-based clustering algorithm and project them in 2D for visualisation.

Let us observe however, that for both of these data exploration strategies 
the two objectives -- clustering and visualisation -- are decoupled and essentially one is entirely subordinated to the other.
This is worrying in that inevitably, the errors accumulated in the first stage cannot be corrected in a subsequent stage and may 
essentially compromise the process of understanding the data. 

In this paper we consider the class visualisation problem, as in \cite{PE} and we identify a setting where a more fruitful coupling between clustering and visualisation 
can be achieved by integrating them both into a consistent probabilistic model. As we shall see, this is particularly useful in high-dimensional problems, where
the number of data dimensions exceeds the number of observations. Such cases are encountered in modern scientific data analysis, e.g. in gene expression analysis, or 
the analysis of special objects in astronomy. 
 
Our approach is based on a multi-objective formulation in a probabilistic formalism. 
As we shall see, our model 
can be decoupled and understood as a sum of two objectives: One
term is responsible for clustering and one other for a PE-like projection of the estimated class assignments. 
These two steps are now 
interdependent, so that in high dimensional problems  
the projection step fulfils a regularisation role, 
guarding against the curse of dimensionality problem.

We use both synthetic data and two real-world high-dimensional data
sets in our experiments: Observed
spectra (of rare quality and coverage)
of early-type galaxies and a benchmark gene expression
array data set are used to demonstrate the working of the proposed approach. 
We find that in both cases we obtain not only a
visualisation of the mixture posteriors, but a more predictive mixture
model, as measured by out of sample test likelihood, as a result of
appropriately combining the objectives of clustering and class
projection.

The remainder of the paper is organised as follows: We begin with presenting our probabilistic model in Section 2. The interpretation by which this can be 
understood as a joint model for mixture based clustering and PE-like projection will become evident in Section 3, where the EM \cite{Bishop} methodology is used 
to derive a maximum a posteriori (MAP) estimation algorithm for our model. We then extend this work to take into account additional available knowledge on measurement 
uncertainties for real-world data analysis. Section 4 presents 
the application two very different experiments involving galaxy spectra and
gene expression. The results are summarised in the concluding section.

\section{The model}
Consider $N$ independent, $T$-dimensional data points. The $n$-th point is denoted by $\bd{d}_n \in \RR^{T\times 1}$ having features $d_{tn}$. 
We seek a 2D mapping of this data into points $\bd{x}_n \in \RR^{2\times 1}$ in the Euclidean space such as to reflect topological relationships based on 
some cluster structure that may be present in the data set. 
In building up our model, we begin by making the common assumption of
conditional independence, in order to enforce the dependences among data features to be captured in the latent space.
\begin{equation}
p(\bd{d}_n|\bd{x}_n)=\prod_t p(d_{tn}|\bd{x}_n)
\end{equation}
Further, in order to capture complicated density shapes, including possibly distinct clusters, 
we model the conditional probabilities of the data features as a mixtures of Gaussians. The mixing coefficients of these mixtures are instance-specific, so that
each measurement that belongs to the same object will have the same mixing coefficient. This will ensure that the various features of
an instance are likely to be allocated to 
the same (set of) mixture components. 
\begin{equation}
p(d_{tn}|\bd{x}_n)=\sum_kp_{\theta_{tk}}(d_{tn}|k)P_{\bd{c}_k}(k|\bd{x}_n)
\label{mixt}
\end{equation}
Observe we do not impose that each data point must belong to exactly one mixture component. This allows us to model the relationships between clusters.

Assuming that we work with real-valued observations, and $p(d_{tn}|k)$ is a Gaussian, then $\theta_{tk}=\{\mu_{tk},v_{tk}\}$ are the mean and 
precision parameters respectively.
Other choices are however possible as appropriate.
\begin{equation}
p(d_{tn}|k)=\cc(d_{tn}|\mu_{tk},v_{tk})
\end{equation}

The second factor in (\ref{mixt}) is a nonlinear function that projects a point $\bd{x}_n$ from the Euclidean space onto a probability simplex. A parameterised 
softmax function can be used for this purpose.
\begin{equation}
P_{\bd{c}_k}(k|\bd{x}_n) = \frac{\exp\ny -\frac{1}{2}(\bd{x}_n-\bd{c}_k)^2 \zr}{\sum_{k'}\exp \ny -\frac{1}{2}(\bd{x}_n-\bd{c}_{k'})^2 \zr}
\end{equation}
Our goal is then to determine $\bd{x}_n$ for each $\bd{d}_n$. In addition, we also need to estimate the parameters $\theta_{tk}$ and $\bd{c}_k$.

In order to somewhat narrow down the search space, we add smoothing priors, similarly to \cite{PE}: 
\begin{equation}
\bd{x}_n \sim \cc(\bd{x}_n|\bd{0},\alpha\bd{I}); \mbox{  } \bd{c}_k \sim \cc(\bd{c}_k|\bd{0},\beta\bd{I});
\end{equation}
In addition, the inverse variance parameters (precisions) are given exponential priors to prevent them produce singularities and encourage the extinction of 
unnecessary model complexity. 
\begin{equation}
p(v_{tk})=\gamma e^{-\gamma v_{tk}}
\end{equation}
The hyperparameters $\alpha, \beta$ and $\gamma$ must all have strictly positive values.

\section{Parameter estimation}
Here we derive MAP estimates for our model specified in the previous section. The complete data log likelihood is proportional to the 
posterior over all hidden variables and this is the following.
{\footnotesize
\begin{eqnarray}
{\mathcal L}&=&\sum_{n,t} \log \sum_k p_{\theta_{tn}}(d_{tn}|k)P_{\bd{c}_k}(k|\bd{x}_n)+\sum_k \log P(\bd{c}_k)+\sum_n
\log P(\bd{x}_n)+\sum_{t,k}\log P(v_{tk}) \nonumber\\
\label{ll}
&\geq& \sum_n \sum_t \sum_k r_{ktn} \ny \log p_{\theta_{tk}}(d_{tn}|k)+\log P_{\bd{c}_k}(k|\bd{x}_n)-\log r_{ktn}\zr \nonumber\\
\mbox{   }&+&\sum_n \log P(\bd{x}_n)+\sum_k \log P(\bd{c}_k)+\sum_{t,k}\log P(v_{tk}) 
\label{Q}
\end{eqnarray}
}
where we used Jensen's inequality and $r_{ktn}\geq0, \sum_k r_{ktn}=1$ represent variational parameters that can be obtained from maximising (\ref{Q}):
\begin{equation}
r_{ktn}=\frac{p_{\theta_{tk}}(d_{tn}|k)P_{\bd{c}_k}(k|\bd{x}_n)}{\sum_{k'}p_{\theta_{tk}}(d_{tn}|k)P_{\bd{c}_k}(k|\bd{x}_n)}
\label{r}
\end{equation}
We can also regard $k=1,...,K$ as the outcome of a hidden class variable and $r_{ktn}$ 
are in fact true class posterior probabilities of this class variable, cf. Bayes' rule.  

The re-writing (\ref{Q}-\ref{r}) is convenient for deriving the estimation algorithm for the parameters of the model. Before proceeding, 
let us rearrange (\ref{Q}) in two main terms, so that the interpretation of our model as a combination of mixture-based clustering 
and a PE-like class projection becomes evident.
{\footnotesize
\begin{eqnarray}
\mbox{Term}_1 &=& \sum_n \sum_t \sum_k r_{ktn} \log p_{\theta_{tk}}(d_{tn}|k)-\gamma\sum_{t,k}v_{tk}+const.\\
\mbox{Term}_2 &=& \sum_n \sum_t \sum_k r_{ktn} \ny \log P_{\bd{c}_k}(k|\bd{x}_n)-\log r_{ktn} \zr +\sum_k
\log P(\bd{c}_k)+\sum_n \log P(\bd{x}_n) \nonumber\\
&=& \sum_{n,t} -KL(\bd{r}_{.,t,n}||P_{\bd{c}_.}(.|\bd{x}_n)) - \alpha\sum_n ||\bd{x}||^2_n - \beta\sum_k ||\bd{c}||^2_k + const.
\end{eqnarray}
}
Now, the first term can be recognised as a clustering model, essentially an instance of modular mixtures \cite{Attias} or an 
aspect-mixture of Gaussians \cite{Hof,Giro}, which is known to be advantageous in high-dimensional clustering problems \cite{Attias}. 
The second term, in turn, is a PE-like objective \cite{PE}, which minimises the Kullback-Leibler (KL) divergence between the class-posteriors and their projections.
Evidently, these two objectives are now interdependent. 
It remains to be seen in which cases their coupling is advantageous.

Carrying out the optimisation of (\ref{Q}) yields the following maximum likelihood estimates for the means and
maximum a posteriori estimates for the precisions.
\begin{equation}
\mu_{tk}=\frac{\sum_n r_{ktn}d_{tn}}{\sum_n r_{ktn}}; {\mbox{          }}
v_{tk}=\frac{\sum_n r_{ktn}}{\sum_n r_{ktn}(d_{tn}-\mu_{tk})^2+2\gamma}
\label{pars}
\end{equation}

For the remaining parameters, there is no closed form solution, we employ numerical optimisation using the gradients 
(see Appendix):
\begin{eqnarray}
\frac{\partial}{\partial \bd{x}_n}&=& \sum_k (\bd{c}_k-\bd{x}_n)\sum_t (r_{ktn} - P_{\bd{c}_{k}}(k|\bd{x}_n))-\alpha \bd{x}_n
\label{latent1}\\
\frac{\partial}{\partial \bd{c}_k}&=& \sum_n (\bd{x}_n-\bd{c}_k)\sum_t (r_{ktn} - P_{\bd{c}_k}(k|\bd{x}_n)) -\beta\bd{c}_k 
\label{latent2}
\end{eqnarray}

As expected, the form of parameter updates also reflects the interdependence of our two objectives: (\ref{pars}) is formally identical with the updates 
in \cite{Attias,Hof,Giro} (up to the variation implied by the use of the prior for precisions\footnote{
In 
\cite{Giro}, the authors propose an improper prior for the variances. Incidentally, our MAP estimate for the precision
parameter (\ref{pars}) is formally identical to the inverse of their variance estimates -- so we now see the expression can be derived 
with the use of a proper exponential prior on the precisions.}). The gradients (\ref{latent1})-(\ref{latent2}) are in turn, as expected, very similar 
to the updates in PE \cite{PE}.

\subsection{Empirical Bayesian test likelihood}
Having estimated the model, the empirical Bayesian estimate \cite{EB} of the goodness of fit for new points is given by 
integrating over the empirical distribution $\frac{1}{N}\sum_n \delta(\bd{x}-\bd{x}_n)$. This is the following.
\begin{equation}
p(\bd{d}_{test})=\frac{1}{N}\sum_n p(\bd{d}_{test}|\bd{x}_n)
\end{equation}

\subsection{Visualisation}
The 2D coordinates $\bd{x}_n, n=1:N$ provide a visual summary of the data density.
In addition, label markers (or colouring information), to aid the visual analysis, are obtained directly from $P_{\bd{c}_k}(k|\bd{x})$. 
This is a handy feature of our method, as opposed to techniques based on dimensionality reduction methods (such as e.g. PCA), 
where detecting meaningful clusters from the projection plot is not straightforward. The class labels may also serve as an interface to the domain expert, 
who may wish to specify and fix the labels of certain points in order to explore the structure of the data interactively.

For accommodating new data points on a visualisation produced from a training set, 
a fast 'folding in' \cite{Hof} procedure can be used. This is to compute $\argmax{\bd{x}} p(\bd{d}_{test}|\bd{x})$ with 
all model parameters kept fixed to their estimated values. 
Conveniently, this optimisation task is convex, i.e. with all other parameters kept fixed, the Hessian w.r.t.
$\bd{x}$ is positive semidefinite 
{\footnotesize
\begin{equation}
\frac{\partial^2 {\mathcal L}}{\partial \bd{x}_n\partial \bd{x}_n^{T}}=\sum_k \sum_t r_{ktn}\bd{c}_k\bd{c}_k^{T}-\ny\sum_k\sum_t r_{ktn}\bd{c}_k \zr 
\ny \sum_k \sum_t r_{ktn}\bd{c}_k\zr^T
\end{equation}
}
for the same reasons as in the case of PE \cite{PE}. Therefore the projection of test points is unique.
However, PE \cite{PE} makes no mention of how to accommodate new points on an existing visualisation plot.

\subsection{Taking into account estimates of  observational error}

It is often the case that data from science domains (e.g. 
astronomy) come with known observational errors.  In this section we
modify our algorithm to take these into account.  Let $\sigma_{tn}$ be
the standard deviation of the known measurement error of the $t$-th
feature of instance $n$. We handle this by considering $d_{tn}$ as a
hidden variable which stands for the clean data, and in addition we
have $\cc(y_{tn}|d_{tn},1/\sigma_{tn}^2)$.

Assuming that we are dealing with real valued data, $p(d_{tn}|\bd{x}_n)$ was defined as a mixture of Gaussians, and so 
the integration over the unseen clean data variable $d_{tn}$ gives the following likelihood
term for component $k$ of feature $t$:
\begin{equation}
p(y_{tn}|k)=\int d d_{tn} \cc(y_{tn}|d_{tn},1/\sigma_{tn}^2)\cc(d_{tn}|\mu_{tk},v_{tk}) = \cc(y_{tn}|\mu_{tk}, (\sigma_{tn}^2+1/v_{tk})^{-1}) 
\end{equation}
In other words, the variance of the data likelihood now has two terms, one coming from the measurement error and one other coming from the modelling error. The latter needs to be estimated.

The estimation equations in this case modify as follows:
\begin{equation}
r_{ktn}=\frac{\cc(y_{tn}|\mu_{tk},(\sigma_{tn}^2+1/v_{tk})^{-1})P_{\bd{c}_k}(k|\bd{x}_n)}{\sum_{k'}\cc(y_{tn}|\mu_{tk},\sigma_{tn}^2+1/v_{tk})P_{\bd{c}_k}(k|\bd{x}_n)}
\label{r_noise}
\end{equation}
The update equation of $\mu_{tk}$ becomes
\begin{equation}
\mu_{tk}=\frac{\sum_n y_{tn}r_{ktn}/(\sigma_{tn}^2+1/v_{tk})}{\sum_n r_{ktn}/(\sigma_{tn}^2+1/v_{tk})}
\end{equation}
and the updates of $\bd{x}_n$ and $\bd{c}_k$ remain unchanged.

For the precision parameters $v_{tk}$ there is no closed form solution and so numerical optimisation may be employed, e.g. a 
conjugate gradient w.r.t. $\log v_{tk}$, since then the optimisation is unconstrained.  
\begin{equation}
\frac{\partial}{\partial \log v_{tk}}=
\frac{1}{v_{tk}}\sum_n r_{ktn} \ny \frac{1}{2(\sigma_{tn}^2+1/v_{tk})} - \frac{(d_{tn}-\mu_{tk})^2}{2(\sigma_{tn}^2+1/v_{tk})^2}\zr-\gamma v_{tk}=0
\end{equation}
Observe that when $\sigma_{tn}=0$, all equations of this subsection reduce to those presented for the noise-free case in Sec. 3. 
Yet another alternative is to treat $d_{tn}$ as hidden variables
and take a hierarchical EM approach. 

It should be highlighted, that although many non-probabilistic methods simply ignore the measurement errors even when these are known, due to our
probabilistic framework a principled treatment is possible. This prevents finding 'interesting' patterns in the visualisation plot
as a result of measurement errors, at least in the cases when such errors are known.
Furthermore, there are cases when further refinement of the noise model will be needed, e.g. 
in many cases the recorded error values are uncertain or known to be optimistic.

\section{Experiments and Applications}
\subsection{Numerical simulations}
The first set of experiments is meant to demonstrate the working of our method and to highlight in which situations it is advantageous
over the fully disjoint and sequential application of a mixture-based clustering and subsequent visualisation strategy. 
Illustrative cases are shown and these are 
important for knowing in what kind of problems is the method appropriate to use.

Throughout, we used smoothness hyperparameters $\alpha=\beta=1$ and $\gamma$ was determined by 
cross-validation under an initial assumption of a large (K=10) number of clusters. The priors on the precision parameters favour 
the extinction of unnecessary components and even if there are remaining redundant components, a good-enough $\gamma$ parameter 
can be located. The typical value obtained was of the order of $10^{-3}$. Then $\gamma$ is fixed and a further 
cross-validation is run to determine the optimal number of clusters $K$ (less or equal to the number of non-empty clusters found in the previous step). 
We noted the optimisation is very sensitive to initialisation and starting $\bd{\mu}_k$ from K-means is beneficial. 
To alleviate problems with local optima, each run was repeated 20 times and the model 
that found better maximum of the model likelihood was retained for further testing.

Two sets of generated data were created. For the first set, 300 points were drawn from a 6-dimensional mixture of 5
independent Gaussians (60 points in each class). The second set was sampled from a 300-dimensional mixture of 5 
independent Gaussians (again, 60 points per class). Fig.~1.a) shows the test likelihood averaged over 20 repeated runs, 
having generated the test data from the same model as the training data. 
The test likelihood obtained with a mixture of Gaussians (MoG) is 
superimposed for comparison. We see that for the relatively low dimensional data set 
the proposed joint model has little (no) advantage. This is simply because in this case there is enough data to reliably estimate a MoG. 
The obtained mixture posteriors could then safely be fed into e.g. a PE \cite{PE} for class visualisation. 

For the case of high dimensional data, however the situation is different. The MoG overfits 
badly and is therefore unable to 
identify the clusters or to provide reliable input to a subsequent visualisation method. This is the situation when our proposed 
model is of use. The 
projection part of the objective guards against overfitting -- as we can see from the test likelihood on Fig~1.b. 
\begin{figure} [here]
\includegraphics[width=2.99cm,height=4.0cm]{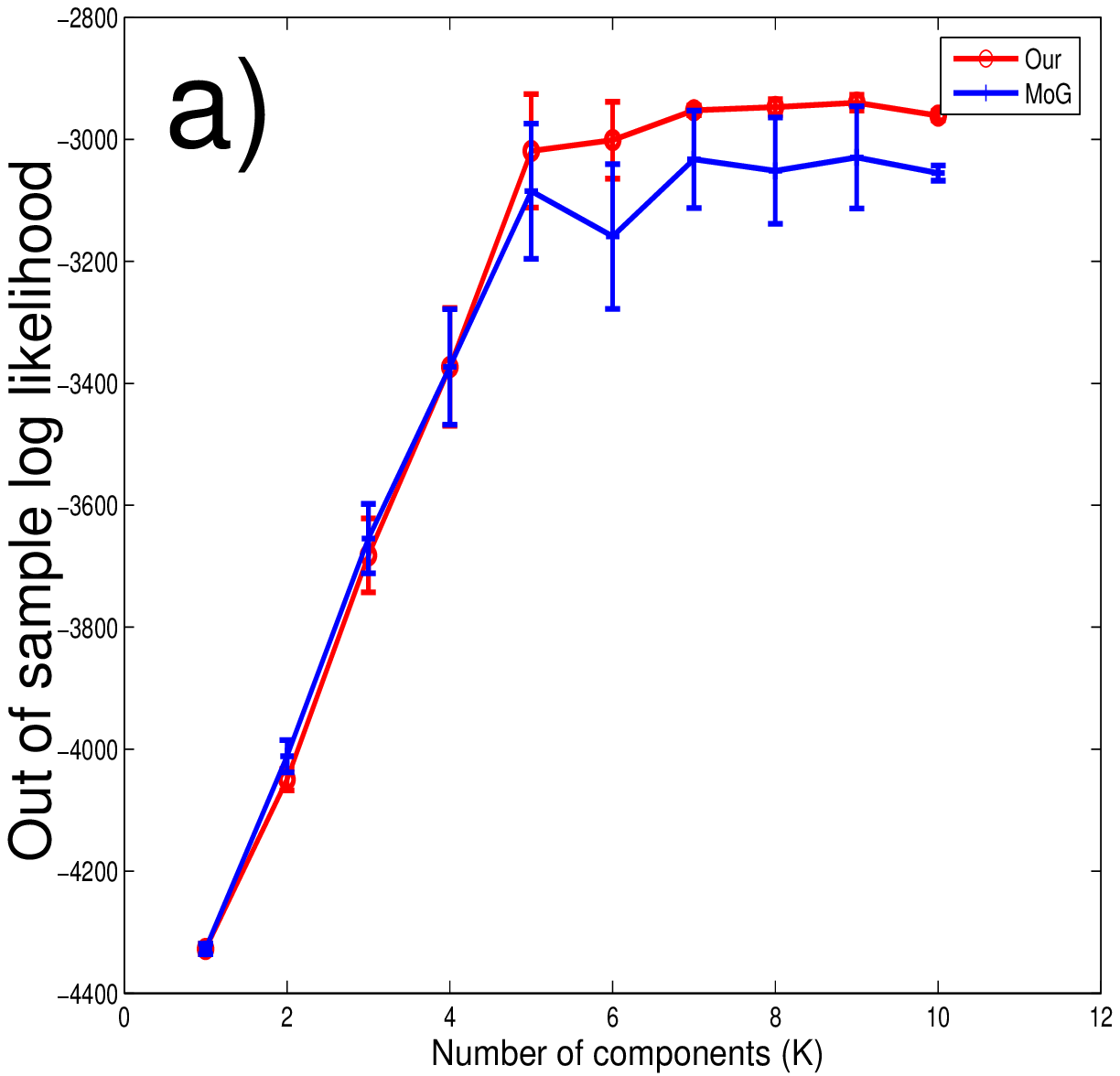}
\includegraphics[width=2.99cm,height=4.0cm]{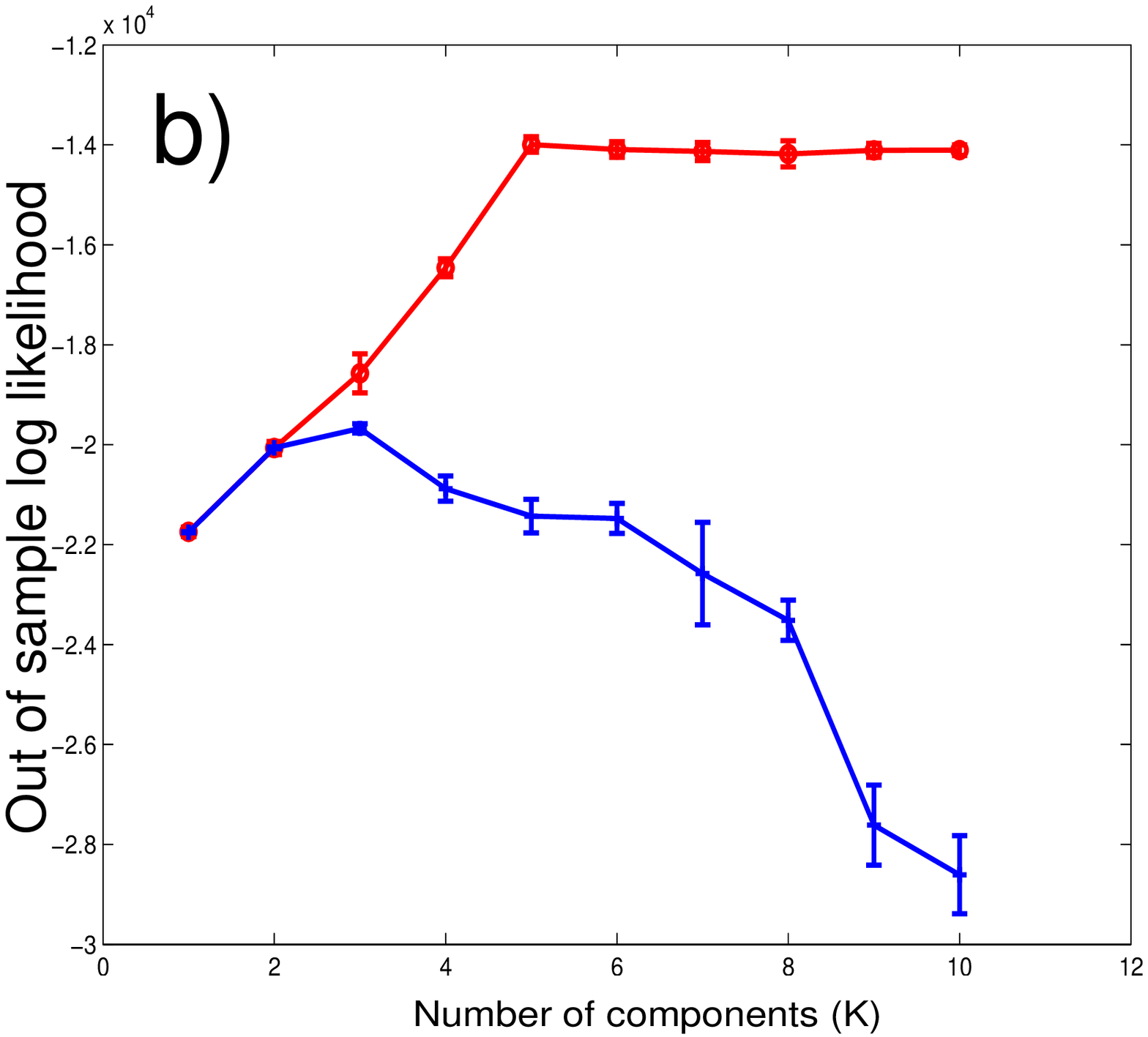}
\includegraphics[width=2.99cm,height=4.0cm]{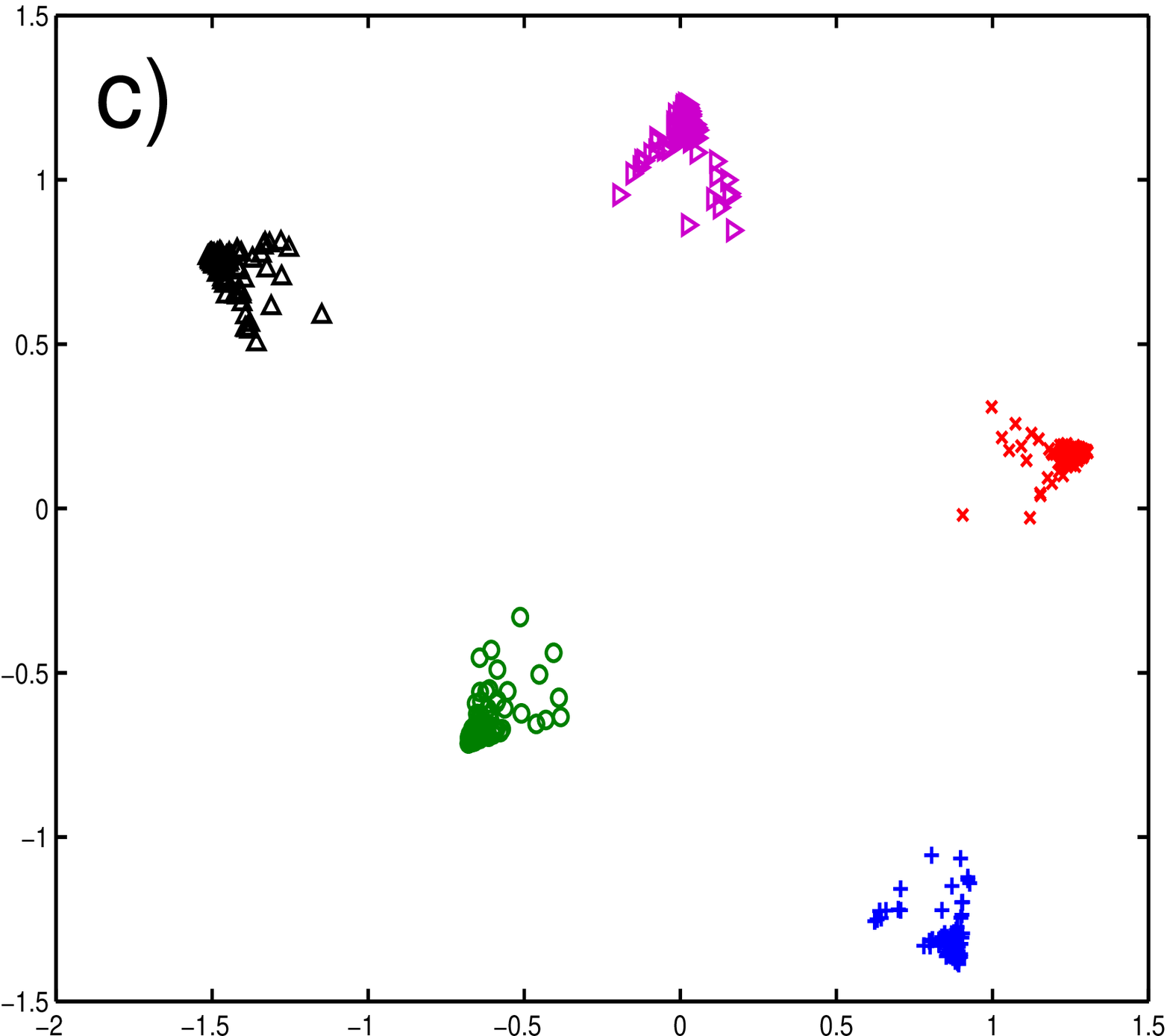}
\includegraphics[width=2.99cm,height=4.0cm]{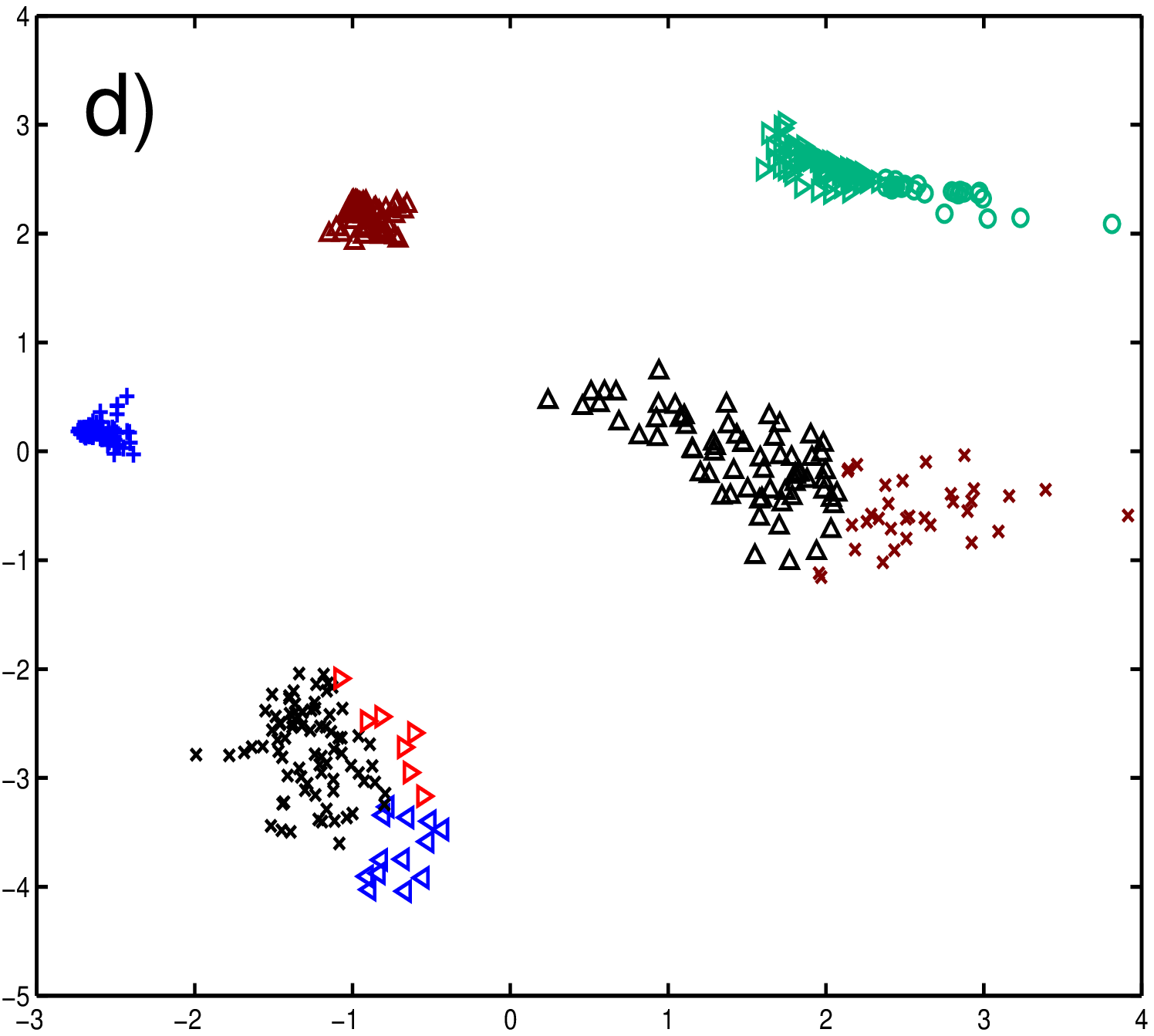}
\caption{
Experiments on synthetic data: a) Out of sample test likelihood (higher is better) for 
the 6-dimensional data --- our approach vs MoG. 
b) Same for the 300-dimensional data. Our approach is significantly less prone to overfitting than MoG
in this case. c) The final visualisation plot for the 300-dimensional data set. 
The markers are automatically assigned by the model and in this case they are identical with the true generator classes.
d) Illustration of the visualisation process when the number of assumed clusters (ten) is larger than the number of true clusters (five).
Notice the true clusters remain compact in the visualisation space.
}
\label{1}
\end{figure}

Fig~1.c shows the visualisation of the 300-dimensional data set. 
Each point is the 2D representation ($\bd{x}_n$) of the corresponding 300-D datum point.
The markers correspond to the maximum argument of the softmax outputs $P(k|\bd{x}_n)$, so they represent labels
automatically assigned by the model. In this case, the estimated labels are identical with the true labels. 
We also see that the true number of classes has 
been correctly recovered. In addition, we noted that in experiments where the number of classes was deliberately chosen 
larger than the true 
number of clusters, some of the unnecessary clusters progressively become empty indeed, due to the employed prior on the precision parameters,   
while others split a true cluster. However, notably, the visual image of the true cluster split by several components tends to remain compact.
Such an example is seen on Fig~1.d.

\subsection{Visualisation of observed spectra of early-type galaxies}

We apply the method developed above to a sample of 
measured spectra (in the
ultraviolet to optical range of radiation) of 21
well-studied early-type (elliptical or lenticular) galaxies.  In a
previous work \cite{sdm,Nolan}, we had studied this data set using various
factor analysis techniques. Here, we seek to obtain a visual analysis
of the data.  Each of these spectra represent flux measurements at 348
values of wavelength in the range 2000-8000\AA, in equal bins,
for all spectra. Observational errors are associated with each value,
which we take into account as described in Section 3.3.  Thus, the
clustering and class visualisation of these spectra is a
high-dimensional problem.

This represents a pilot data set for an important study in the
evolution of galaxies. It is generally believed that all early-type
galaxies formed all their stars in the early universe, and then have
evolved more-or-less passively until the present day- so one expects
to find their spectrum to correspond to a collection of stars all of
the same age. However, detailed observations in the last decade
indicate a wealth of complex detail in a significant fraction of such
galaxies, including evidence of a sub-population of very young stars
in many cases.  How common this effect is largely unknown, and can
only be addressed through data mining of large spectral archives. Even
though many $\times 10^5$ galaxy spectra are being assembled in large
public archives (e.g. www.sdss.org), a sample as detailed as ours is
rare and difficult to assemble, particularly with such wide a coverage
in wavelength, which requires combining observations from both ground
and space based observatories (see details in \cite{Nolan}).  From
this small sample, we would attempt to isolate those galaxies which
have young stars from those that don't. 

Needless to say, the fluxes are all positive values. In order to be
interpretable, our method needs to ensure the estimated parameters
(cluster prototypes) are also positive. In our previous work, we built
in explicit constraints to ensure this \cite{Nolan}.  Here, since each
$\mu_{tk}$ in (\ref{pars}) is just a weighted average of positive data,
its positivity is automatically satisfied.

The leave-one-out test likelihood of our model
is shown on the left plot of Fig.~2. The peak at $K=2$
indicates that two clusters have been identified.  A mixture of
Gaussians test likelihood is superimposed for comparison, showing the
overfitting problem due to the high dimensionality.  The MoG is
therefore unable to identify any clusters in the data. Hence, a class
visualisation of the data based on mixture posteriors would be clearly
compromised in this case. The right hand plot shows the grouping of
the actual data around the two identified prototypes $\bd{\mu}_k,
k=1,2$ of our model.  The latter are superimposed with thick
lines. These can be recognised and interpreted as the prototype
of the spectrum of a  `young'
and  `mature' stellar population respectively. 
Thus, in this case,  the clusters have a
physical interpretation in astrophysical terms.

\begin{figure} [here!]
\begin{center}
\includegraphics[width=5.6cm,height=4.5cm]{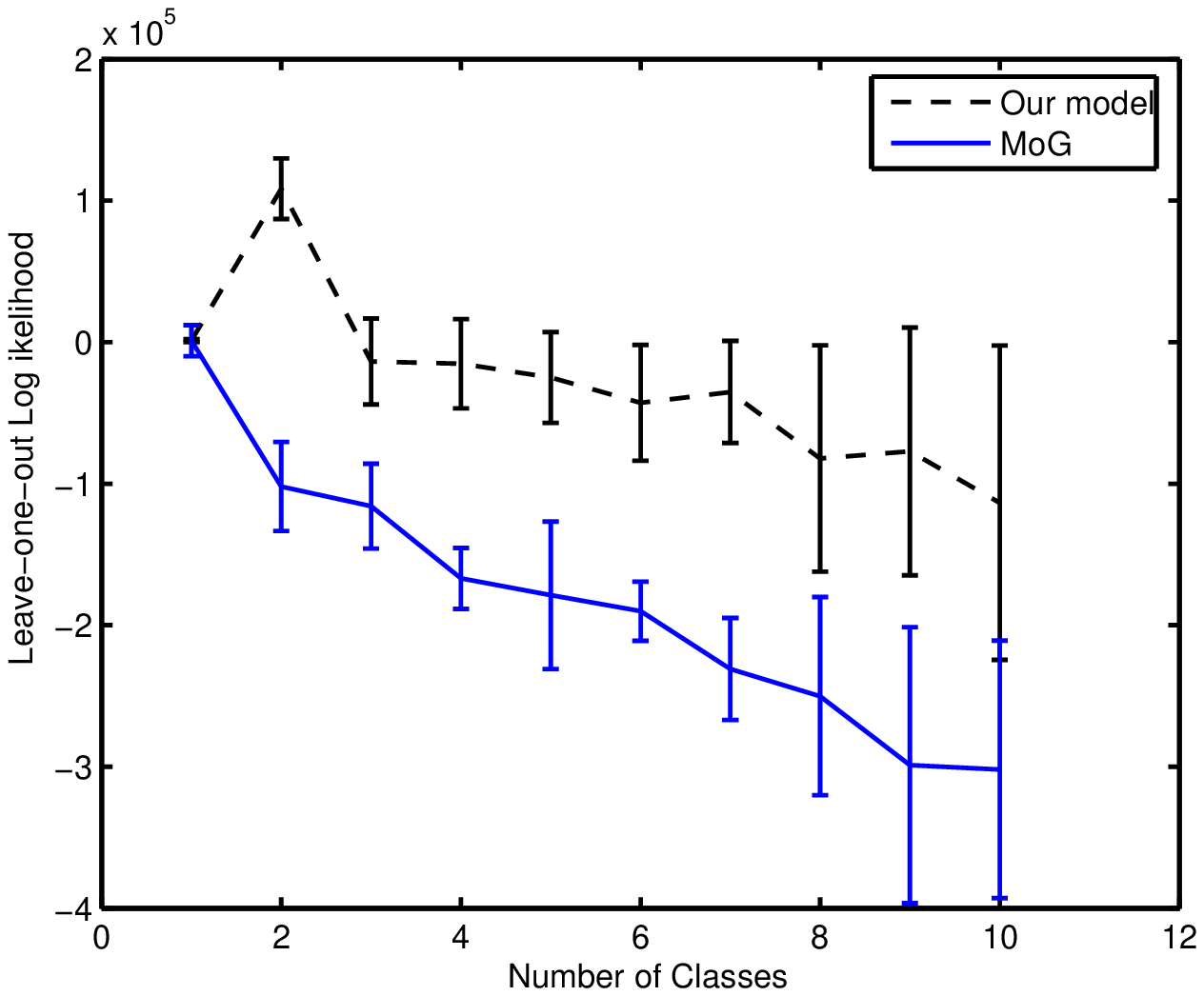}
\includegraphics[width=5.8cm,height=4.5cm]{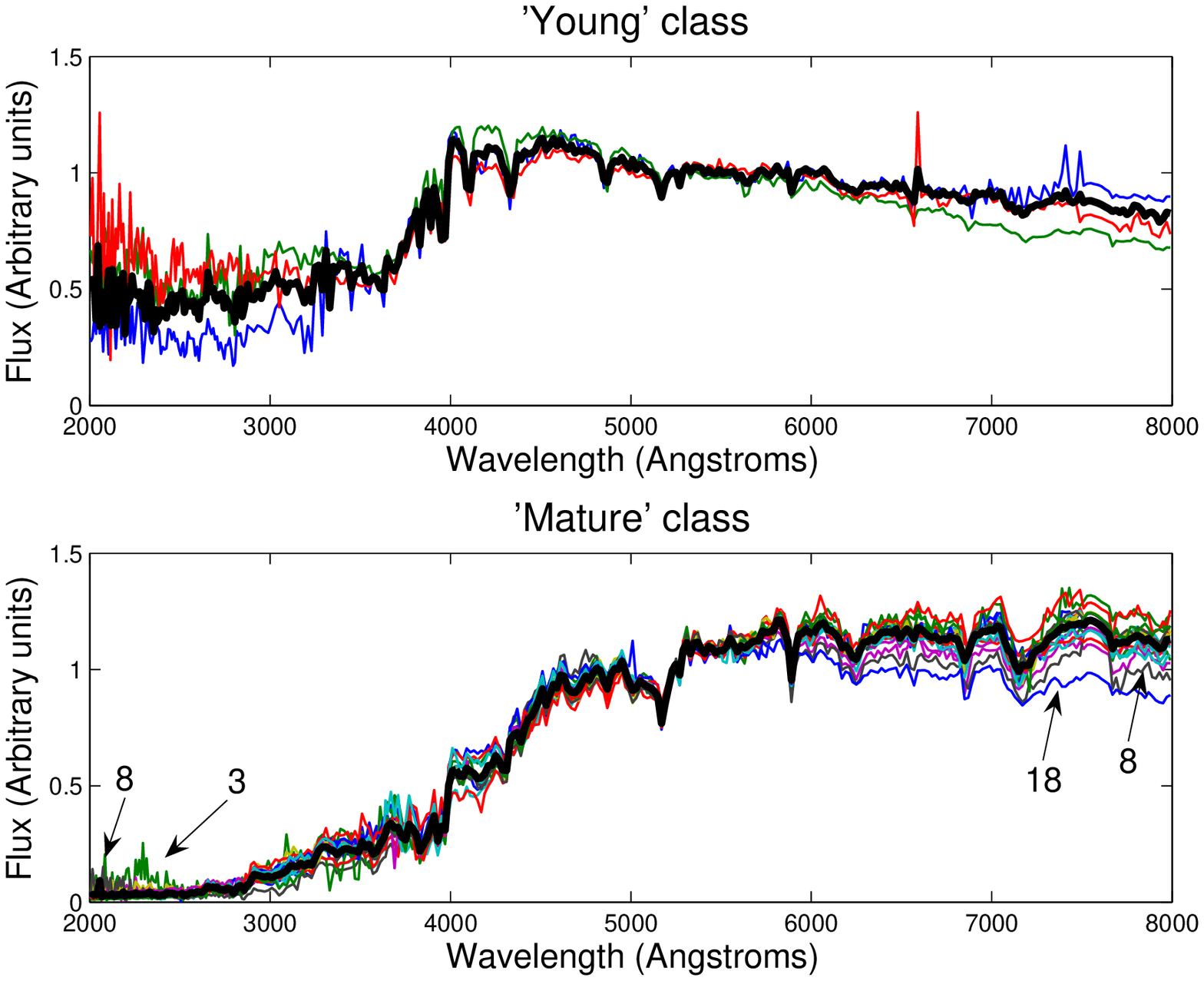}
\caption{
Left: The leave-one-out test likelihood versus the number of clusters. 
The peak at $K=2$ produced by our method indicates two clusters whereas a mixture of Gaussians overfits 
and therefore fails to identify clusters in the data. 
Right: The actual data, clustered around the two prototypes identified by our model. 
The parameters $\bd{\mu}_k$ for $k=1,2$ are superimposed with thick lines. They are interpretable 
as a 'young' and an 'old' prototypical spectrum respectively. The identification number that marks some of the spectra correspond to
those on Fig.~3. The marked spectra are the instances that apart from their overall shape present some features of 
the 'young' category too.
}
\end{center}
\end{figure}

Identification numbers mark some of the spectra clustered in the
`mature' category on the left lower plot of Fig.~2.  These are the
galaxies that have a significantly non-zero class membership for
either cluster, and they indeed include some morphological
aspects of the `young' category as well (the emission lines at
$<2000$\AA\  and the slope of the spectral continuum
in the range 6000-8000\AA). Physically, this indicates the presence
of a significant population of young ($<2$~Gyr old) stars, whereas the rest 
of the stars are $>10$~Gyr old. 

\begin{figure} [here!]
\begin{center}
\includegraphics[width=9.5cm,height=4.5cm]{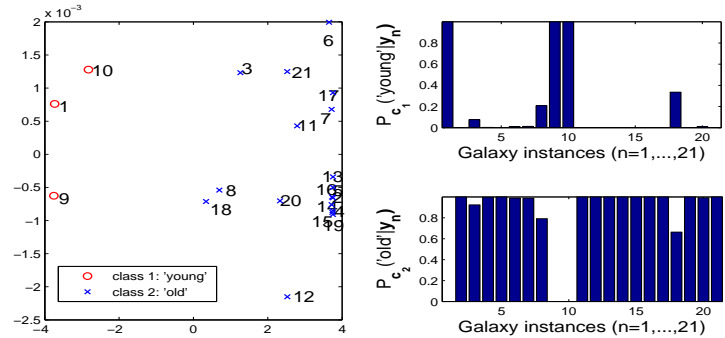}
\caption{
Left: Visualisation of the data set of the spectra of 21 early-type galaxies. For each spectrum $\bd{y}_n$, 
the 2D latent representation $\bd{x}_n$ is plotted. The markers are those assigned by the model, as shown on
the right.}
\label{astro}
\end{center}
\end{figure}

The identification numbers are the same as those on Fig.~3, where we see
the 2D visualisation of the sample on the left.  For each spectrum
$\bd{y}_n$, the 2D latent representation $\bd{x}_n$ is plotted. The
markers represent cluster labels 
automatically assigned by the model,
as detailed in the right hand plot. We see the two clusters are well
separated on the image and the `hybrid' galaxies are indeed placed in
between those that are clearly cluster members.  
Of these, the one marked as 18 represents a galaxy (NGC 3605) for which recent detailed 
physical analyses have been made \cite{Nolan_new}. It turns out that although more than 85\% of its stellar mass
is associated with an old (9--12 Gyr) stellar population, it does contain a younger stellar population too,
at $\simeq 1$ Gyr \cite{Nolan_new}. 

We therefore conclude that it is possible to have an
intuitive visual summary of a
few hundreds of measurements per galaxy in just two coordinates
with the application of our method.

\subsection{Visual analysis of gene expressions}
In a brief final experiment we show the potential use of our approach for the visual
analysis of high dimensional gene expression arrays.  Oligonucleotide
arrays can provide a means of studying the state of a cell, by
monitoring the expression level of thousands of genes at the same time
\cite{Alon} and have been the focus of extensive research.  
The main difficulty is that the number of
examples is typically of the order of tens while the number of genes
is of the order of thousands. Even after eliminating genes that have
little variation, we are still left with at least hundreds of data
dimensions. Straightforward mixture based clustering runs into the
well-know curse of dimensionality problem.  
Here we apply our method
to the ColonCancer data set, having 40 tumour and 22 normal colon
tissue samples \cite{Alon}.  This is a benchmark data set, used in
many previous classification and clustering studies.  
Our input matrix consisted of the 500 genes with highest overall variation $\times$ the 62 samples.

\begin{figure} [here]
\begin{center}
\includegraphics[width=6.3cm,height=4.5cm]{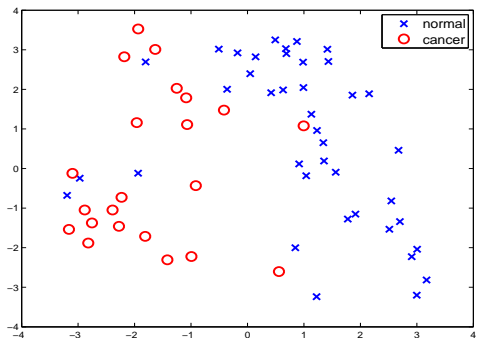}
\caption{\footnotesize
Unsupervised class visualisation of the ColonCancer data set. 
The markers correspond to the true class for the ease of visual evaluation.}
\label{colon}
\end{center}
\end{figure}

Fig.~\ref{colon} shows the visualisation obtained in a purely unsupervised manner. The markers now correspond 
to the true labels (not used by the algorithm), and are given for the ease of visual evaluation of the representation produced. 
The separation of cancerous from noncancerous tissues is most apparent on the plot. The potential of 
such a visualisation tool lies mainly in that it would allow a domain expert to interactively explore
the structure of the sample. Additionally, the gene-specific posteriors $r_{ktn}$ provide quantitative gene-level class information. 

\section{Conclusions}
We proposed and investigated a model for class visualisation of explicitly high dimensional data. We have shown 
this model relates closely to PE \cite{PE} in that it represents a probabilistic integration of the clustering and visualisation 
objectives into a single model. We derived empirical Bayesian estimates for our model which make this multi-objective interpretation 
easy to follow. 
Although this work may potentially further be enhanced by a fuller Bayesian estimation scheme, the empirical Bayesian methodology
has been appropriate for our purposes \cite{EB_astro} and it allows us to estimate an empirical latent density from a given 
example set of data and to reason about previously unseen data relative to that.  
We demonstrated gains in terms of the predictive capabilities of the proposed model 
over the fully modular and sequential approach to clustering and class visualisation in the case of high dimensional data. 

\subsubsection*{Acknowledgements}
This research is funded by PPARC grant PP/C503138/1, `Designer Algorithms for Astronomical Data Mining'. AK also acknowledges
partial support from a Wellcome Trust VIP Award (Project 10835). 

\footnotesize
\section*{Appendix. Estimation of $\bd{x}_n$}
The terms containing $\bd{x}_n$ are the following.
\begin{eqnarray*}
& & \sum_n \sum_t \sum_k r_{ktn} \ny -\frac{1}{2}(\bd{x}_n-\bd{c}_k)^2 - \log \sum_{k'} \exp(-\frac{1}{2}(\bd{x}_n - \bd{c}_k)^2)\zr - \alpha \bd{x}_n^2 \\
&=& \sum_n \sum_t\ny -\sum_k r_{ktn} \frac{1}{2}(\bd{x}_n-\bd{c}_k)^2 - \log \sum _{k'} \exp(-\frac{1}{2}(\bd{x}_n - \bd{c}_k)^2) \zr -\alpha \bd{x}_n^2
\end{eqnarray*}
The gradient is then:
\begin{equation}
\frac{\partial}{\partial \bd{x}_n}=\sum_t \ny -\sum_k r_{ktn}(\bd{x}_n - \bd{c}_k) + \sum_{k''} \frac{\exp(-\frac{1}{2}(\bd{x}_n-\bd{c}_{k''})^2)} 
{\sum_{k'} \exp(-\frac{1}{2}(\bd{x}_n-\bd{c}_{k'})^2)}(\bd{x}_n-\bd{c}_{k''})\zr - \alpha \bd{x}_n \nonumber
\end{equation}
Renaming $k''$ by $k$ and replacing the expression of $P_{\bd{c}_k}(k|\bd{x}_n)$
the following is obtained.
\begin{eqnarray*}
\frac{\partial}{\partial \bd{x}_n}
&=& \sum_t \sum_k \ny -r_{ktn}(\bd{x}_n - \bd{c}_k) + P_{\bd{c}_k}(k|\bd{x}_n)(\bd{x}_n - \bd{c}_k) \zr -\alpha \bd{x}_n\\
&=&\sum_k (\bd{c}_k-\bd{x}_n)\sum_t (r_{ktn} - P_{\bd{c}_{k}}(k|\bd{x}_n))-\alpha \bd{x}_n
\end{eqnarray*}

\end{document}